\documentclass{article}

\usepackage{arxiv}

\usepackage[T1]{fontenc}
\usepackage{placeins}
\usepackage{xcolor}%

\usepackage{amsmath,amsfonts,bm}

\def\eqref#1{equation~\ref{#1}}
\def\1{\bm{1}}

\DeclareMathAlphabet{\mathsfit}{\encodingdefault}{\sfdefault}{m}{sl}
\SetMathAlphabet{\mathsfit}{bold}{\encodingdefault}{\sfdefault}{bx}{n}

\usepackage{graphicx}

\usepackage{subcaption}

\usepackage{hyperref}
\usepackage{url}

\usepackage{rotating} 

\AtBeginDocument{%
  }

\title{Bag of concepts can enhance other text representations for medical systematic review screening}

\author{
 Radosław Pytlak \\
  Faculty of Mathematics and Information Science\\
  Warsaw University of Technology\\
  Koszykowa 75, Warsaw, 00-662\\ Poland \\ \texttt{radoslaw.pytlak@pw.edu.pl}\\
   \And
 Paweł Cichosz \\
  Faculty of Electronics and Information Technology\\
  Warsaw University of Technology\\ Nowowiejska 15/19, Warsaw, 00-665\\ Poland\\ \texttt{pawel.cichosz@pw.edu.pl}\\
  \And
 Bartłomiej Fajdek \\
  Faculty of Mechatronics\\
  Warsaw University of Technology\\ św. Andrzeja Boboli 8, Warsaw, 02-525\\ Poland\\
\texttt{bartlomiej.fajdek@pw.edu.pl} \\
  \And
 Bogdan Jastrzębski \\
Faculty of Mathematics and Information Science\\
  Warsaw University of Technology\\
  Koszykowa 75, Warsaw, 00-662\\ Poland \\ \texttt{bogdan.jastrzebski.DOKT@pw.edu.pl}\\
}

\begin{document}

\maketitle

\begin{abstract}
The paper examines the choice of text representation for semi-automated systematic literature review (SLR) systems based on active learning. Motivated by the insufficient attention that the bag of concepts representation has received so far, the paper discusses the implementation and experimental evaluation of several variations thereof. The experimental study was performed on medical SLR case studies, with results reported for random forest classification, cluster initialization and uncertainty sampling. The utility of bag of concepts features obtained using different ontologies is evaluated by applying them within the active learning framework to sequentially create classification models based on training data iteratively expanded with expert annotations. Systematic experiments compare BoW, BoC, PubMedBERT, Jina, Gemma and Qwen3 representations, together with combinations of BoC and each of the other representations. The evaluation uses workload savings and performance profiles adapted from optimization benchmarking. The results suggest that concept-based features can provide complementary information, with the BoC-augmented representations achieving higher mean workload savings on the DERP cases than their corresponding base representations. The Qwen3+BoC representation gives the best average performance in this comparison. The effects of augmentation vary across case studies and base representations. The paper also concentrates on the impact of ontology choice on document classification outcomes. In particular, the paper compares the predictive value of bag of concepts representations based on chosen ontologies with that of representations obtained using MedCAT with a UMLS-based dictionary.

\end{abstract}
\section{Introduction}

The publication of new medical research results requires comparing them with all relevant previously published results; that is, it requires a systematic review of the literature (SLR). 
An SLR is a process that requires the involvement of a number of experts in the field it covers and involves reviewing a large number of articles. If we use the calculations provided in \cite{bib4} and assume that the SLR is performed on the basis of article abstracts, an experienced researcher can review on average 120 brief documents per hour. 
If we assume that an SLR requires reviewing $7,000$ abstracts, the process will require up to 56 hours of expert work (7 person-days). Moreover, the evaluation of abstracts of articles on difficult topics may require much more than two minutes, and when we take into account that at the end of the SLR process there is also a full--text evaluation of a selected small subset of articles, the given process time should be multiplied.

Evidence-based medicine (EBM) has been described by David Sackett as the “conscientious, explicit, and judicious use of current best evidence in making decisions about individual care.” (preface in \cite{bib2}) This means that SLRs are essential tools for EBM. 

The Institute of Medicine, established in 1970 by the National Academy of Sciences, convened the Roundtable in 2006. The Roundtable developed a vision for a learning healthcare system that “draws upon the best evidence to provide the care most appropriate to each patient, emphasizes prevention and health promotion, delivers the most value, adds to learning throughout the delivery of care, and leads to improvements in the nation’s health.” The Roundtable set the goal that, by 2020, $90$ percent of clinical decisions would be evidence-based (preface in \cite{bib2}).

Due to the importance of SLRs in EBM, several applications based on machine learning techniques were introduced in order to speed up SLRs (\cite{t3}, \cite{t6}, \cite{t8}, \cite{t13}, \cite{t14}, \cite{cheng2018colandr}, \cite{simon2019bioreader}, \cite{t10}, 
\cite{t11}, \cite{t12}). These applications use the active learning process in document classification and usually use bag of words (BoW) as a text representation.  While more recent neural network-based embeddings have become popular in several natural language processing applications \cite{hasny2023bert}, they do not necessarily outperform BoW in text classification tasks \cite{graff2025bag}.

The paper examines another approach for creating a text representation, bag of concepts (BoC), which has a strong resemblance to BoW but is focused on the use of domain information contained in ontologies. We present a systematic evaluation of the usefulness of BoC for document classification, both separately and in combination with lexical and neural representations, using various concept vocabularies within an active learning framework. In this work, we demonstrate how the choice of ontology can influence classification results using the most commonly used ontologies in medicine. Because the BoC representation requires the use of dictionary-based annotators, we present classification results using reputable annotators. The evaluation concerns screening based on article abstracts.

This paper builds on the BoC framework and evaluation approach reported in \cite{Pytlak_2025}, which compared BoW, BoC and their combination on 15 DERP case studies. That study also compared classifiers, methods for selecting the initial training set, and query selection strategies. The active learning configuration employed here is based on the results of those comparisons. The main extension examined in the present paper is the systematic evaluation of the augmentation of text representations, including not only BoW but also deep learning-based embeddings, with BoC features. The study also compares representations obtained using different ontology combinations and the NOBLE and MedCAT annotators.

\section{Methods}

The work we present is based on standard machine learning techniques to build classification models in an active learning process, using BoW, BoC and neural text representations and their combinations. The paper concentrates on the description of our BoC approach to text representation and pays less attention to other details of the implementation of the active learning process. In particular, we present results obtained with the random forest (RF) classifier, selected on the basis of the comparisons with SVM and naive Bayes reported in \cite{Pytlak_2025}.

%%%%---
\subsection{Bag of concepts}
\label{subsec:boc}
%%%%---
The BoC representation is essentially BoW with concepts used instead of words. 
Our approach to BoC is similar to that described in \cite{bib13}, but with one crucial difference. The approach in \cite{bib13} relies on the UMLS system and its ontology management tools. In our approach, to obtain BoC representations of articles, the following steps must be performed: 1) choosing ontologies for annotation vocabularies; 2) annotating articles by using a dictionary-based annotator; 3) determining the BoC representation of articles; 4) creating extended versions of the BoC representations by referring to ontology properties.  

The extended BoC representation includes concepts related to the annotated concepts through their positions in the ontology trees used to build the annotator dictionary (see also \cite{pytlakBIBM2023},\cite{Pytlak_2025}). We used the ROBOT application \cite{bib21} for ontology management. In particular, it was used to determine the paths to the root of the ontology tree from its selected vertices that represent concepts. In order to correctly determine the paths, it was necessary to determine all the relationships between the concepts. In order to accomplish this task, Semantic Reasoner was employed.

Our procedure for building the BoC representation requires a dictionary-based annotator. In our work we used the annotators NOBLE and MedCAT---the first one was chosen on the basis of several criteria such as reported accuracy, time efficiency, and ease of incorporation into our SLR application (the review paper \cite{bib26} persuaded us to incorporate the NOBLE annotator into our system; the other annotators we took into account were the cTAKES annotator \cite{bib24}, and the Neji annotator \cite{bib25}). The MedCAT annotator was chosen due to its reported accuracy and unique functionality for resolving concept ambiguity (\cite{MedCAT21}).

In the case of the NOBLE annotator, when we applied it with more than one ontology dictionary, we annotated texts sequentially with different dictionaries. It must be admitted that this approach implies that terms identified by an annotator can appear in more than one ontology. 
%%%%---
\subsubsection{The application of the MedCAT annotator} 
%%%%---
MedCAT (Medical Concept Annotation Toolkit) was built with the purpose of providing a tool for facilitating analysis of medical records kept by medical institutions (\cite{MedCAT21}). It is a dictionary-based annotator equipped with ambiguity-resolving functionality. In order to determine the meaning of a concept found in an annotated text, we need an embedding of the concept and the embeddings of words (tokens) surrounding the concept. Since MedCAT stores embeddings of concepts (at least some of them) which were constructed on the basis of word embeddings and then updated according to the embeddings of their contexts (by using unsupervised learning exploiting a large corpus of medical texts), we can evaluate the embedding of a concept in question, update it according to its context and then compare it to stored embeddings. In that way we can resolve the concept ambiguity. The information about dictionary concepts (including their embeddings) is stored in MedCAT's CDB (Concept Database).

The CDB we created was based on the UMLS Metathesaurus which we downloaded from \cite{UMLS} to obtain 2 708 631 concepts and 5 722 139 synonyms. We trained the UMLS MedCAT model on the data set composed of abstracts from PubMed and full texts from PMC (as available in September 2013) on the basis of the word embeddings provided by the NLPLab initiative (\cite{BioNLP}).
The BioNLPLab embeddings were trained on the collection of biomedical abstracts and full texts using the word2vec algorithm. It is reported in \cite{MedCAT21} that using other popular models for building concept embeddings such as fastText, GloVe and BERT did not improve the performance of the MedCAT annotator. 

Training the UMLS MedCAT model produced embeddings for 318 830 concepts, which constitute around $11.8\%$ of all concepts. After creating the UMLS MedCAT model we compared the performance of the annotator based on it to the performance of the MedCAT annotator described in \cite{MedCAT21}. In order to evaluate the performance of both implementations of the UMLS models we used them to annotate texts in the set MedMentions introduced in \cite{MedMen}. MedMentions is the set of medical texts which were manually annotated by experts on the basis of the concepts present in the UMLS Metathesaurus. Our model achieved the score $F_1 =0.442$ while the score of the annotator from \cite{MedCAT21} was $F_1 =0.448$. Our UMLS MedCAT model thus achieved annotation performance close to that of the reference model on MedMentions.

\subsubsection{Feature space} 
The basic element used in ontology creation is a taxonomy tree. Concepts located closer to the root of the tree are more general than those located further away. The location of a concept's node in the ontology tree can be used in its categorization. Similar to \cite{bib13}, we introduced a concept depth threshold that determines which concepts should be used in the extended version of the BoC representation. In order to determine concept depths, we used the reasoner to generate the ontology tree which contains not only edges of the taxonomy tree but also edges corresponding to relations between concepts revealed by the reasoner. Following \cite{bib13}, we determined the concept depth on the basis of the shortest path of the concept node to the root of the ontology tree.

The process of building the BoC representation of articles from the set $P$ begins with determining the set of concepts from a given ontology that appear in at least one article $p$ from $P$. The set $V$ obtained in this way is called the \emph{concept vocabulary}. Then, for each concept $v$ from the set $V$ we determine the number of articles $N(v)$ in which this concept occurs. If we denote by $n(p,v)$ the number of occurrences of the concept $v\in V$ in the article $p\in P$, then the value of the feature assigned to the concept $v$ in the BoC representation of the article $p$---$f(p,v)$---is calculated as follows:
\begin{eqnarray}
  % f(p,v) & = & n(p,v)\cdot d(v), \label{TF}\\
  % f(p,v) & = & \left (1+ \log n(p,v) \right )\cdot
  %           \log\frac{N}{N(v)}\cdot d(v), \label{TF_IDF}
  f(p,v) & = & n(p,v), \label{TF}\\
  f(p,v) & = & \left (1+ \log n(p,v) \right )\cdot
            \log\frac{N}{N(v)}, \label{TF_IDF}
\end{eqnarray}
where $N$ is the number of articles. (\ref{TF}) corresponds to the TF representation and (\ref{TF_IDF}) to the TF-IDF variant for $n(p,v)>0$; the feature value is zero when the concept does not occur in the article. Notice that in the case of the TF-IDF formula the sublinear transformation of the term frequency was used to avoid assigning too much weight to multiple concept occurrences in the same article. Due to the enormous number of nodes in the UMLS Metathesaurus taxonomy tree the depth coefficient $d(v)$ was set to $1$ in the BoC representation obtained with the help of the MedCAT annotator (we were not able to evaluate the shortest paths for the nodes of the UMLS tree). 

In our work, we also used another concept vocabulary---the extended concept vocabulary in which the concept vocabulary obtained as described above  is extended by certain concepts whose vertices are on the shortest paths drawn from the vertices of the concepts found by the annotator to the root of the ontology tree. 
Suppose that $v\in V$. For concept $v$, we create a set $EV(v)$ such that $v\in EV(v)$ and all other concepts $\tilde{v}$ whose vertices lie on the shortest path from vertex $v$ to the root of the ontology tree and which have depth $d(\tilde{v}) \geq TL$ are also included, where $TL$ is a fixed depth threshold level. Then we create an enlarged concept vocabulary: $EV = \bigcup_{v\in V}EV(v)$ and modify the method of calculating feature values for individual concepts, according to the formula:
\begin{align}
% f(p,\tilde{v}) ={} & v_r \times f(p,v),\ \  v_r =\frac{1}{d(v)-d(\tilde{v}) + 1},\ v\in V,\ \tilde{v}\in EV(v),
f(p,\tilde{v}) ={} & v_r \times f(p,v),\ \  v_r =\frac{d(v)}{d(v)-d(\tilde{v}) + 1},\ v\in V,\ \tilde{v}\in EV(v),\nonumber 
\end{align}
where $d(v)$ is the depth of the concept $v$. This way of creating a concept vocabulary was proposed in the paper \cite{bib13}, which also presents an illustration of the creation of this concept vocabulary (see also \cite{Pytlak_2025}).

\subsection{BERT--based representation}
\label{subsec: bert}

BERT (Bidirectional Encoder Representations from Transformers) is a family of
deep language models based on the transformer architecture and the
self-attention mechanism. During pre--training, selected input tokens are masked
and predicted using information from both their left and right contexts. This
produces context-dependent representations in which the representation of a
word depends on the surrounding text \cite{devlin19bert}. Sentence--BERT--type
models adapt this architecture to produce a single, semantically meaningful
embedding vector for an entire text sequence \cite{reimers2019sentencebert}.

PubMedBERT retains the BERT architecture but is pre-trained from scratch
exclusively on biomedical literature, using a vocabulary derived from the same
domain. Its training corpora comprise PubMed abstracts and PubMed Central
full-text articles \cite{gu2022pubmedbert}. This domain-specific pre-training
is intended to provide more adequate representations of biomedical terminology
than models trained primarily on general-domain text. PubMedBERT has previously
been successfully applied to biomedical article classification \cite{fang2022bioformer}. Fixed PubMedBERT text embeddings have also been used
as inputs to conventional classification algorithms
\cite{kondadadi2026pubmedbert}.

The pre--trained model used in this work to obtain article representations is
\texttt{NeuML/pubmedbert-base-embeddings} \cite{mezzetti2023pubmedbert}.
It is a PubMedBERT-base model adapted within
the Sentence Transformers framework \cite{reimers2019sentencebert} using
PubMed title--abstract pairs and pairs of similar titles. Each
complete article abstract is supplied to the model as a single text sequence
and directly mapped to one 768-dimensional dense embedding vector. The model
accepts sequences of at most 512 subword tokens; when this limit is exceeded,
it automatically truncates the input. The
resulting vectors are L2--normalized, so that every embedding has unit length.
The normalized sequence-level representation
returned by the model is used as the article's fixed attribute vector.

\subsection{LLM--based representations}
\label{subsec: llms}

In addition to the domain--specific PubMedBERT model, we employ three embedding
models derived from large language models to obtain article representations.

Jina Embeddings v5 is a family of compact text embedding models developed by
Jina AI~\cite{akram2026jinaembeddingsv5} (which we will subsequently refer to as the Jina model). The models are trained using a novel
regimen that combines knowledge distillation from larger teacher models with
task-specific contrastive learning, producing compact yet high-performance
embeddings. The \texttt{jina-embeddings-v5-text-small} variant is used in this
work. It supports input sequences of up to 32\,768 tokens across many
languages and maps each article to a 1024-dimensional dense embedding vector.
The model has achieved state-of-the-art results for models of comparable size
on standard retrieval, clustering, and classification benchmarks.

Qwen3--Embedding is a family of text embedding models built upon the Qwen3
foundation models~\cite{zhang2025qwen3embedding} (referred to below as Qwen3). Its training pipeline combines large--scale unsupervised pre--training with supervised fine-tuning on
high--quality data synthesized by the Qwen3 language models themselves. The
8--billion--parameter variant is used in this work. It supports input sequences
of up to 32\,768 tokens and produces embedding vectors of up to 4096
dimensions; the full 4096-dimensional output is used here. The model has
achieved state-of-the-art results on the MTEB multilingual benchmark and
supports more than 100 languages.

EmbeddingGemma is a 300--million--parameter open embedding model developed by
Google, built from the Gemma~3 architecture with a T5Gemma
initialization~\cite{gemmateam2025gemma3} (referred to below as Gemma). It was trained on approximately 320-~billion tokens of web documents, code and technical documents, and
synthetic task-specific data spanning more than 100 languages. Despite its
compact size, the model is designed for resource-constrained deployment while
retaining competitive embedding quality. Each article abstract is supplied as
a single text sequence and mapped to one 768-dimensional dense embedding
vector. The model accepts input sequences of at most 2048 tokens.

All three models are accessed through the Ollama inference server and used to obtain fixed text embeddings for a classifier.
%and applied
%to the concatenation of each article's title and abstract. 
The resulting
fixed-length vectors are used directly as article attribute representations,
providing a complementary set of embeddings that differ from the PubMedBERT
representations in model scale, training data composition, and architectural
design.

\subsection{Active learning}
\label{subsec:al}

Active learning is a machine learning scheme in which the learning system actively participates in the formation of its training set. In the most common pool-based setting, this is performed by selecting some instances from an unlabeled dataset and querying a human annotator or some other knowledge source to provide their labels \cite{cohn94active}. This is well suited to the SLR process where human experts are supposed to annotate query articles as relevant (included) or irrelevant (excluded).

As the training set is expanded during active learning, a sequence of classification models is obtained, with the first one created based on a small initial training set and subsequent ones created on bigger training sets obtained by adding labeled query articles.

In the paper \cite{Pytlak_2025}, several active learning strategies were investigated in the context of supporting systematic literature reviews. Regarding the initial selection of the training set, the following schemes were considered: a) selecting the initial training set at random; b) selecting the most representative initial training articles from cluster centers and cluster borders \cite{yuan11initial}. In the cluster method, cluster membership is determined using the fuzzy c-means algorithm \cite{bezdek81fuzzy}. Article vectors are transformed to a lower-dimensional space using latent semantic analysis \cite{dumais05lsa}. 

At each subsequent iteration of active learning, articles are typically selected based on the predictions of the model created in the previous iteration, but other factors such as article similarities may be used as well \cite{fu13selection}. The following methods were used in \cite{Pytlak_2025}: 1) selecting articles for which the predictions of the current model have the highest uncertainty \cite{nguyen22uncertainty}; 2) selecting articles with the highest dissimilarity to those already present in the training set; 3) selecting articles from the regions of the highest density; 4) selecting articles at random (used as a comparison baseline).

Simulation experiments conducted and reported in \cite{Pytlak_2025} demonstrated that an active learning strategy employing variant b) for the creation of the initial training set and variant 1) for the query selection yielded the best classification results. Although these findings were obtained using the BoW and BoC representations, they provide an empirical rationale for reusing cluster initialization and uncertainty sampling in the present study. We therefore retain this configuration to compare the different text representations under a common active learning setup.

\subsection{Performance evaluation}
\label{subsec:eval}

Classification performance is evaluated in terms of article screening workload savings to assess the utility of the models for the intended SLR application. Multiple setups of the learning process, using different article representations, are compared with respect to their performance profiles, similar to those used for optimization solvers \cite{DM_2002}. 

\subsection{Workload savings}
\label{subsec:wss}

SLR workload savings due to using a classification model can be estimated based on the \textit{Yield} and \textit{Burden} of the process \cite{bib4}:
%\begin{align}
\begin{eqnarray}
{\displaystyle   \textit{Yield} = \frac{\textrm{TP}^L+\textrm{TP}^U}
          {\textrm{TP}^L+\textrm{TP}^U+\textrm{FN}^U},\ \ \textit{Burden} =
    \frac{\textrm{TP}^L+\textrm{TN}^L+\textrm{TP}^U+\textrm{FP}^U}{N}}\nonumber
%\end{align}
\end{eqnarray}
where \textrm{TP} is the number of relevant articles predicted to be so by the model,
\textrm{TN} is the number of irrelevant articles predicted to be so by the model, \textrm{FP} is the number of irrelevant articles predicted to be relevant by the model, \textrm{FN} is the number of relevant articles predicted to be irrelevant, and $N$ is the number of all articles. When these quantities are determined based on labeled articles from the training set, the $L$ superscript is used, and the $U$ superscript is used for quantities corresponding to unlabeled articles. Actual model predictions are only used for unlabeled articles, and for labeled articles their known true classes are used as `predictions' for the purpose of this evaluation.

Workload savings defined using \textit{Yield} and \textit{Burden} represent the amount of human screening work saved by using model predictions while achieving a required level of \textit{Yield}, e.g., $0.95$ \cite{bib3}. This leads to the following \textit{WSS@95\%} metric (work saved for random sampling at $95$\%):
\begin{equation}
  \textit{WSS@95\%} = 1-\textit{Burden} - (1-\textit{Yield}) \;\;|\;\; \textit{Yield}\geq 0.95.
\end{equation}

When evaluating model performance, all possible decision thresholds in the $[0, 1]$ interval for the relevant class probability are considered to identify the maximum threshold which provides \textit{Yield} of at least $0.95$.
This threshold value is used to determine predicted classes for \textit{WSS@95\%} calculation. We then identify the active learning iteration at which the maximum \textit{WSS@95\%} is achieved, along with the percentage of manually screened articles to which this iteration corresponds. Both selections are made retrospectively using the known labels of the complete dataset, including articles treated as unlabeled during learning. The reported metric values characterize workload savings under this retrospective evaluation procedure. The reported percentages identify the selected training iterations; the total screening burden also includes the remaining articles predicted as relevant.

\subsection{Performance profiles}

In order to determine an effective strategy for conducting active learning in document classification, one should consider different classifiers, different text embeddings, different methods of determining the initial set of documents for manual review by experts, etc. Furthermore, these arrangements must be evaluated on many sets of documents. 

Discussing original experimental data that is the result of considering many variants of the learning process may lead to erroneous conclusions.
As argued in \cite{DM_2002}, using data aggregation, applying medians and quantiles of the selected metric, or ranking lists can only partially solve the problem of the complexity of data to be compared. Following \cite{DM_2002}, we apply the methodology of optimization performance profiles to the analysis of active learning arrangements. 

Suppose that we have a set of classification strategies $\mathcal{S}$ and we want to evaluate them on sets of documents $\mathcal{D}$, on the basis of \textit{WSS@95\%}. Then for any strategy $s\in\mathcal{S}$ and any set of documents $d\in\mathcal{D}$ we denote by $w_{s,d}$ the value of \textit{WSS@95\%} obtained by the strategy $s$ applied to the set  $d$. 
In order to determine the performance profile of the selected strategy, we first calculate the coefficient $r_{s,d}$ representing the relative performance of strategy $s$ applied to the set of documents $d$ in relation to the highest performance obtained on the set $d$ by any strategy from the set $\mathcal{S}$
\begin{equation}
    r_{s,d} = \frac{w_{s,d}}{\max_{s'\in\mathcal{S}} w_{s',d}},\ \ {\rm as\ a\ result}\ \    p_s(\tau) = \frac{1}{|\mathcal{D}|} \big|\{d\in\mathcal{D}: r_{s,d} \geq \tau \}\big|
\end{equation}
%is treated as the probability of the result of strategy $s$ being  within the factor of $\tau$ of the best recorded result. 
The function $p_s(\tau)$ is the fraction of datasets on which strategy $s$ achieves at least $\tau$ times the best recorded result for the specified comparator set---the higher the curve the better the relative performance.
Functions $p_s:\ [0,1]\rightarrow [0,1]$ are non-increasing and values $p_s(1)$, $s\in {\mathcal S}$ indicate the fraction of datasets on which each strategy achieves or ties the best result. These profiles summarize relative performance across equally weighted case studies. Their interpretation refers to the datasets and comparator set $\mathcal{S}$ used in each comparison. 

Performance profiles provide an easily interpretable measure for comparing different active learning process variants: if the performance profile of a given variant lies above the profile of another variant, the former is superior to the latter.

\section{Results}

We experimentally evaluate the usefulness of text representations for active learning in systematic literature reviews on real medical SLR case studies. The reported comparisons use the RF classifier with cluster initialization and uncertainty query selection. This version of the active learning process is the result of an extensive comparison of other variants (utilizing BoW and BoC text representations, naive Bayes, linear SVM and random forest classifiers), the findings of which are presented in \cite{Pytlak_2025}.

\subsection{Data}
\label{subsec:data}

In our comparisons we used 15 SLR case studies created by the Drug Effectiveness Review Project team (DERP)---\cite{bib20}. The SLR case studies were available from \cite{bib9}, and were used in several projects discussing semi--automated systems for SLRs, i.e., \cite{bib10}, \cite{bib16}, \cite{bib13}---these case studies are characterized in Table \ref{table1}. 

\begin{tiny}
\begin{table}
	\centering
	{\caption{Test cases.}\label{table1}}
	\begin{tabular}{lll}
		\hline
		& \textbf{Total} & \textbf{Relevant (\%)} \\ \hline
		ACE Inhibitors & 2544 & 183 (7.19 \%) \\ 
		ADHD & 851 & 84 (9.87 \%) \\ 
		Antihistamines & 310 & 92 (29.68 \%) \\ 
		Atypical Antipsychotics & 1120 & 363 (32.41 \%) \\ 
		Beta Blockers & 2072 & 302 (14.58 \%) \\ 
		Calcium Channel Blockers & 1218 & 279 (22.91 \%) \\ 
		Estrogens & 368 & 80 (21.74 \%) \\ 
		NSAIDS & 393 & 88 (22.39 \%) \\ 
		Opioids & 1915 & 48 (2.51 \%) \\ 
		Oral Hypoglycemics & 503 & 139 (27.63 \%) \\ 
		Proton Pump Inhibitors & 1333 & 238 (17.85 \%) \\ 
		Skeletal Muscle Relaxants & 1643 & 34 (2.07 \%) \\ 
		Statins & 3465 & 173 (4.99 \%) \\ 
		Triptans & 671 & 218 (32.49 \%) \\ 
		Urinary Incontinence & 327 & 78 (23.85 \%) \\ \hline
	\end{tabular}
\end{table}
\end{tiny}

\subsection{Implementation and experimental procedure}

We constructed the BoW model using the \texttt{TfidfVectorizer} class from the Python scikit-learn library \cite{sklearn1.1.3}. Vocabulary construction involved removing stopwords and filtering tokens according to their document frequency. Tokens appearing in fewer than $1$\% or more than $95$\% of the documents were excluded.   

For the BoC representation, we developed our own implementation. The pipeline combines the NOBLE annotator with the biomedical ontologies used in the reported comparisons---SNOMED CT, MeSH, NCI Thesaurus, and ChEBI---and uses a depth cut-off of $TL=3$ to construct the extended representation. These settings were selected in accordance with the recommendations in \cite{bib13}. In further experiments, we replaced NOBLE with MedCAT configured to use the UMLS knowledge model.

The reported experiments use the \texttt{RandomForestClassifier} implementation provided by the \texttt{scikit-learn} Python library \cite{sklearn1.1.3}, with mostly default settings. While this may prevent the classifier from achieving the best possible performance, it is more realistic since hyperparameter tuning would hardly be possible in real SLR application conditions given the very limited availability of class labels. The only non-default setting is enabling class balancing weights.

Our own implementation of the active learning process was used, with initialization and query selection strategies described above. The initial training set size was set to $20$. In the case of cluster initialization, $13$ of them were cluster-border instances and $7$ were cluster-center instances. When necessary, additional articles are selected in batches of $5$ until both classes are represented. During each active learning iteration $5$ query articles are selected. Since our intention is to observe and evaluate the screening performance at all stages of the learning process, it continues until the whole pool is exhausted. On each iteration the available true classes for labeled articles and the predicted class probabilities for unlabeled pool articles are combined and evaluated, with the probabilities of known true classes set to $1$. To reduce the variance resulting from the initial training set selection, each active learning process is repeated $10$ times and for each iteration true class labels and model predictions from these $10$ runs are combined by pooling. The reported performance measures are calculated from these pooled predictions. 

\subsection{Evaluation results}

The reported comparisons include six base text representations: bag of words (BoW), bag of concepts (BoC), PubMedBERT, Jina, Gemma and Qwen3. BoC is also concatenated with each of the other five representations, yielding BoW+BoC, PubMedBERT+BoC, Jina+BoC, Gemma+BoC and Qwen3+BoC. In the figures and tables, BoW+C denotes BoW+BoC, while BERT and BERT+BoC denote PubMedBERT and PubMedBERT+BoC, respectively. All results reported here use the RF classifier with cluster initialization and uncertainty sampling. Comparisons of RF with linear support vector machines and multinomial naive Bayes using BoW and BoC representations of abstracts were reported in \cite{Pytlak_2025} and provided the basis for selecting RF for the present comparisons.
We evaluated the utility of bag of concepts for classifying article abstracts using 15 DERP case studies. 

\begin{figure}
%\centerline{\includegraphics[height=3.5in]{ecaif01}}
%\centerline{\includegraphics[height=3.9in]{figures/wss95_rf_boc_cu_LaT.pdf}}
\centerline{\includegraphics[height=2.5in]{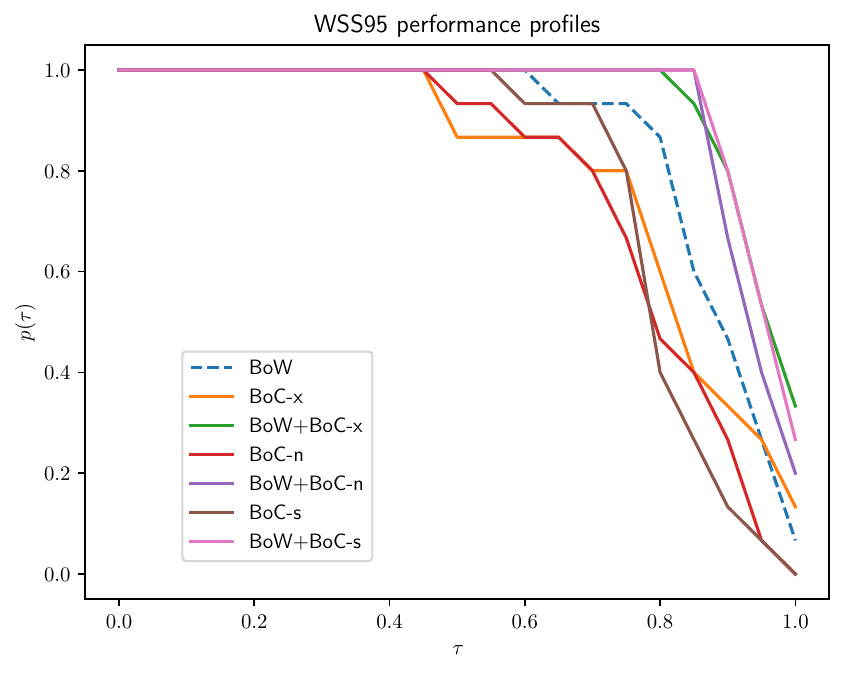}}
% \caption{Performance profiles on DERP for BoW and different BoC/BoC+W variants: normal (n), short (s) and extended (x) for active learning with SNOMED CT--MeSH vocabularies.}
\caption{Performance profiles on DERP for BoW and different BoC/BoW+BoC variants: normal (n), short (s) and extended (x) for active learning with SNOMED CT--MeSH vocabularies.} 
%% \caption{Performance profiles for.} 
\label{wss95_rf_boc_cu}
\end{figure}

The following alternatives for constructing the concept-based vocabulary differ in their expansion and filtering strategies:
\begin{enumerate}
\item normal (n)---incorporates every concept detected by the annotator without any additional filtering or hierarchy--based augmentation.
\item short (s)---identical to the normal setup, but prunes concepts appearing in fewer than $1\%$ or more than $95\%$ of the documents.
\item extended (x)---enlarges the vocabulary by including ancestor concepts on a shortest path from each annotated term to the root, subject to the depth threshold described in Section~\ref{subsec:boc}.
\item extended short (xs) — similar to the extended variant, but concepts 
appearing in fewer than $1\%$ or more than $95\%$ of the documents are excluded.
\end{enumerate}
Figure~\ref{wss95_rf_boc_cu} compares the normal (n), short (s) and extended (x) variants of BoC and BoW+BoC, with BoW as the baseline. The profiles do not show a consistent advantage of vocabulary expansion across the tested cases. The other presented results therefore use the short representation. Table \ref{tableNfeat} presents the number of features used in the short representations of abstracts from 15 case studies of the DERP collection for BoW and BoC representations. Regarding the PubMed-BERT, Jina, Gemma, and Qwen3 representations, the number of features is 768, 1024, 768, and 4096, respectively.

%%%%%%%%%%%%%%%%%%%%%%%%%%%%%%%%%%
\begin{table}
    %\color{red}
	\centering
	\caption{The number of features in the BoW, BoC (s) representations (S+M--SNOMED CT + MeSH, C+N--CheBI + NCI Thhesaurus, Med--MedCAT).}\label{tableNfeat}
	\begin{tabular}{lllll}
		\hline
		& BoW & BoC--S+M &  BoC--C+N & BoC--Med\\ \hline
		ACE Inhibitors & 1121 & 1380 & 2096 & 1223 \\ 
		ADHD & 1132 & 1319 & 2246 & 1460 \\ 
		Antihistamines & 1231 & 1331 & 2727 & 1801 \\ 
		Atypical Antipsychotics & 1088 & 1124 & 1998 & 1173 \\ 
		Beta Blockers & 1187 & 1565 & 2274 & 1272 \\ 
		Calcium Channel Blockers & 1080 & 1327 & 2150 & 1206 \\ 
		Estrogens & 1204 & 1110 & 2510 & 1206 \\ 
		NSAIDS & 1156 & 1353 & 2300 & 1430 \\ 
		Opioids & 1181 & 1230 & 2035 & 1252 \\ 
		Oral Hypoglycemics & 1189 & 1145 & 2176 &  1348\\ 
		Proton Pump Inhibitors & 1021 & 1032 & 1764 & 1128 \\ 
		Skeletal Muscle Relaxants & 1236 & 1794 & 2351 & 1398 \\ 
		Statins & 1201 & 1559 & 2444 & 1435 \\ 
		Triptans & 1156 & 1131 & 1986 & 1240 \\ 
		Urinary Incontinence & 1397 & 1423 & 2727 & 1873 \\ \hline
	\end{tabular}
\end{table}
%%%%%%%%%%%%%%%%%%%%%%%%%%%%%%%%%%%%%%%%%%%%%%%%%

To provide a direct quantitative assessment of the impact of concept augmentation, Table~\ref{paired_summary} presents the mean WSS@95\% values obtained for each base representation and its corresponding version augmented via the BoC method. In addition to the mean differences between pairs of results, the table includes data on the number of cases analysed in which the augmentation improved, left selection performance unchanged, or degraded it. 

\begin{table}[!htbp]
%\color{blue}
\centering
\caption{Paired comparison of mean WSS@95\% for base
and BoC-augmented text representations.}
\label{paired_summary}
\small
\setlength{\tabcolsep}{4pt}
\begin{tabular}{llcccc}
\hline
%Benchmark & Base model & Base & +BoC
Base model & Base & +BoC
& $\Delta$WSS & Better/Tied/Worse \\
\hline
 BoW        & 0.2677 & 0.2814 & +0.0138 & 10/0/5 \\
 PubMedBERT & 0.3026 & 0.3125 & +0.0098 & 10/0/5 \\
 Jina       & 0.3033 & 0.3178 & +0.0145 & 11/0/4 \\
 Gemma      & 0.2917 & 0.3071 & +0.0154 & 11/0/4 \\
 Qwen3      & 0.3165 & 0.3417 & +0.0252 & 10/0/5 \\
%\hline
%SWIFT (5) & BoW        & 0.6205 & 0.6184 & -0.0021 & 2/0/3 \\
%          & PubMedBERT & 0.6440 & 0.6520 & +0.0080 & 2/0/3 \\
\hline
\end{tabular}
\end{table}

Table \ref{paired_summary} summarizes the impact of augmenting base text representations with BoC features. For the DERP dataset, all five augmented representations achieve higher average WSS@95\% values than their corresponding base representations. The greatest average improvement was observed for the Qwen3 representation (+0.0252), followed by Gemma (+0.0154) and Jina (+0.0145). However, this improvement is not uniform across all the analysed systematic reviews. Depending on the base representation, adding BoC features improves results in 10 or 11 of the 15 DERP test cases.

\begin{figure}[htbp]
  \centering
  % --- lewy rysunek ---
  \hspace*{-0.3cm}
  \begin{subfigure}[t]{0.52\textwidth}
    \centering
    \includegraphics[width=\linewidth]{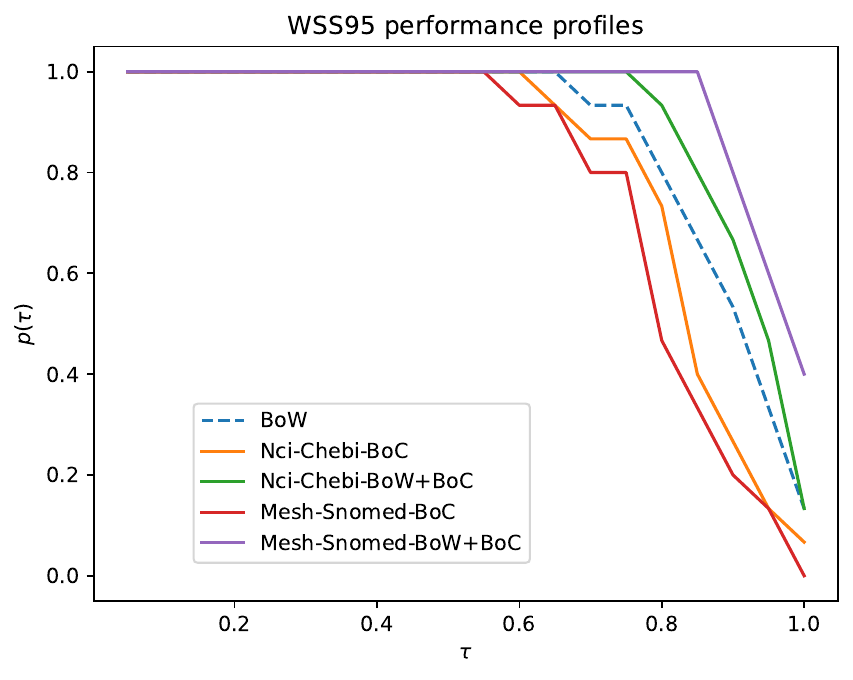}
    % \caption{MeSH--SNOMED CT, ChEBI--NCI.}
    \caption{NOBLE}
    \label{fig:ont1}
  \end{subfigure}%
  % --- prawy rysunek ---
  \begin{subfigure}[t]{0.52\textwidth}
    \centering
    \includegraphics[width=\linewidth]{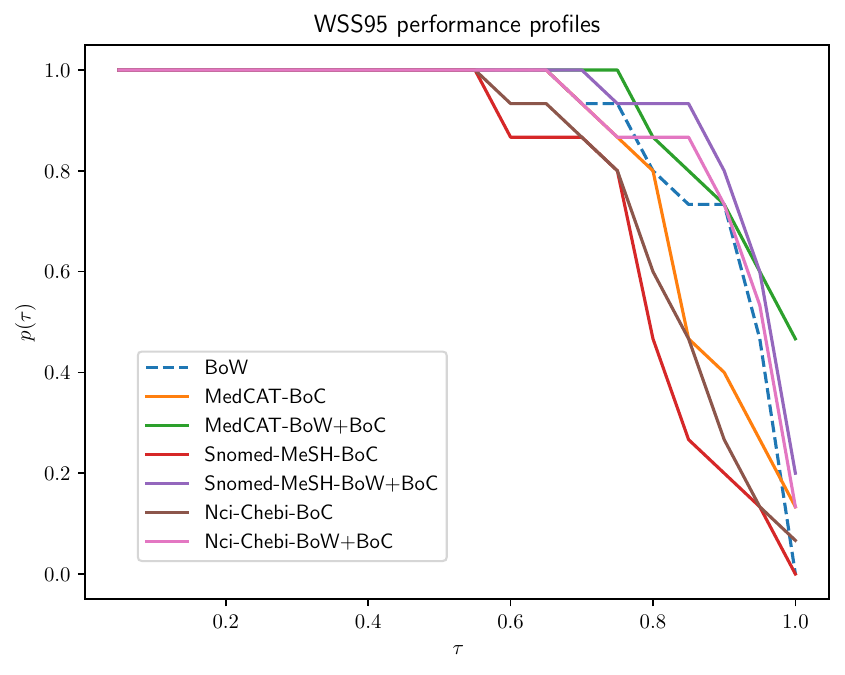}
    % \caption{MeSH--SNOMED CT, ChEBI--NCI, UMLS Metathesaurus (MedCAT).}
    \caption{NOBLE and MedCAT}
    \label{fig:ont2}
  \end{subfigure}
  % \caption{Active learning performance profiles on DERP for BoW and different BoC/BoC+W representations obtained by using various ontology combinations and different annotators: a) NOBLE, b) NOBLE and MedCAT.}
  \caption{DERP performance profiles for BoW, BoC and BoW+BoC. NOBLE uses MeSH--SNOMED CT or ChEBI--NCI; MedCAT uses the UMLS Metathesaurus.}
  \label{fig:active_learning_profiles}
\end{figure}

\begin{figure}
%\centerline{\includegraphics[height=3.5in]{ecaif01}}
%\centerline{\includegraphics[height=3.9in]{figures/wss95_rf_boc_cu_LaT.pdf}}
\centerline{\includegraphics[height=2.5in]{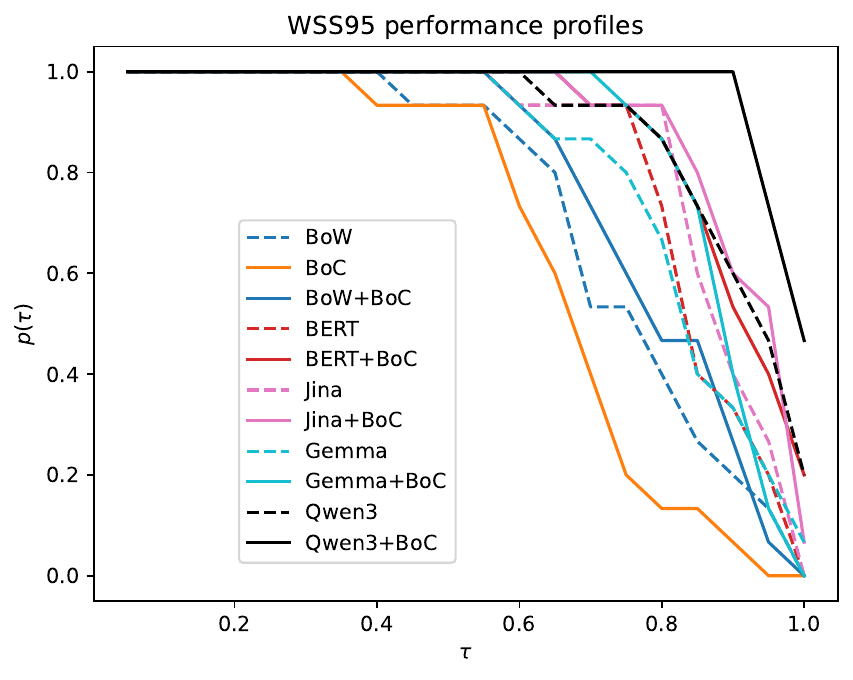}}
% \caption{Performance profiles on DERP for BoW, BoC ( based on SNOMED--CT + MeSH ontologies), BoW, PubMedBERT, Jina, Qwen3 and Gemma representations, including their BoC-augmented variants---for active learning with cluster initialization and uncertainty sampling query selection, and the RF algorithm.} 
\caption{DERP performance profiles for lexical, concept and neural representations, including BoC augmentation. BoC uses MeSH--SNOMED CT vocabularies; BERT denotes PubMedBERT.} 
%% \caption{Performance profiles for.} 
\label{fig:derp_all}
\end{figure}

\begin{figure}[htbp]
  \centering
  % --- lewy rysunek ---
  \hspace*{-0.3cm}
  \begin{subfigure}[t]{0.52\textwidth}
    \centering
\includegraphics[width=\linewidth]{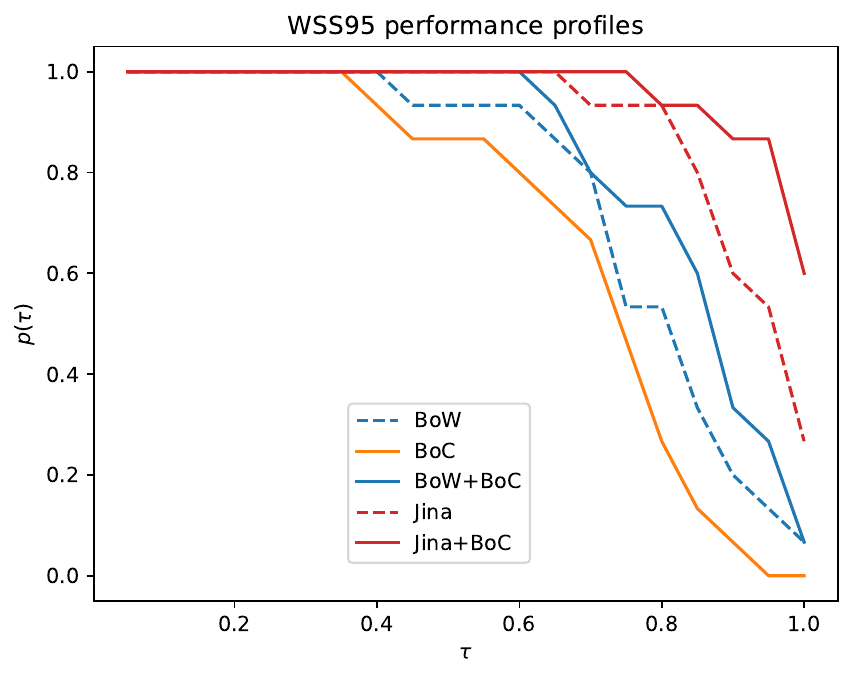}
    % \caption{BoW, BoC, BoW+BoC, Jina, Jina+BoC}
    \caption{Jina}
    \label{fig:derp_jina}
  \end{subfigure}%
  % --- prawy rysunek ---
  \begin{subfigure}[t]{0.52\textwidth}
    \centering
\includegraphics[width=\linewidth]{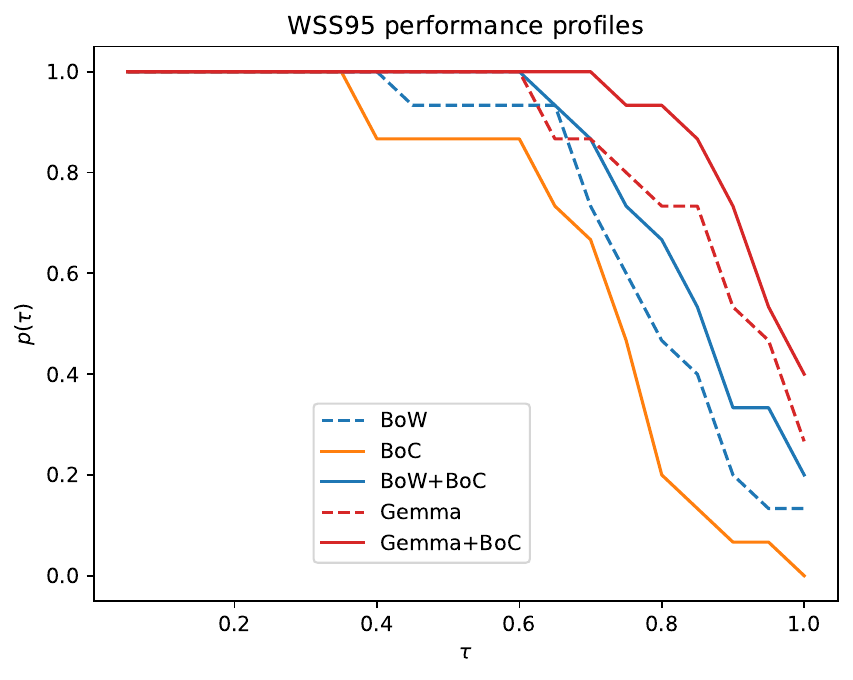}
    % \caption{BoW, BoC, BoW+BoC, Gemma, Gemma+BoC).}
    \caption{Gemma}
    \label{fig:derp_gemma}
  \end{subfigure}
  % \caption{Performance profiles on DERP based on SNOMED--CT + MeSH ontologies.}
  \caption{DERP performance profiles for neural representations with and without BoC augmentation. BoC uses MeSH--SNOMED CT vocabularies.}
  \label{fig:derp_jina_gemma}
\end{figure}

\begin{figure}[htbp]
  \centering
  % --- lewy rysunek ---
  \hspace*{-0.3cm}
  \begin{subfigure}[t]{0.52\textwidth}
    \centering
\includegraphics[width=\linewidth]{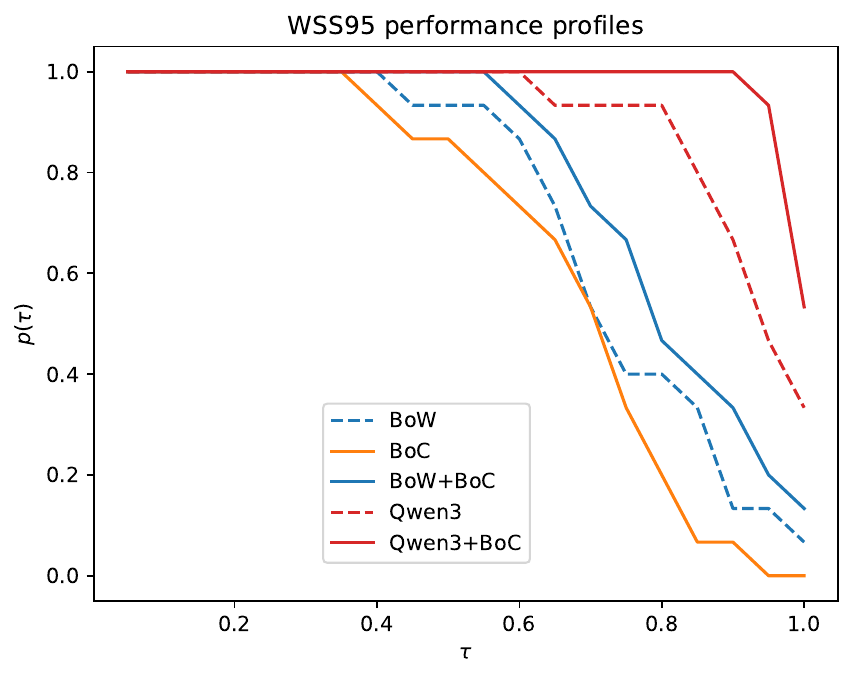}
    % \caption{BoW, BoC, BoW+BoC, Qwen3, Qwen3+BoC}
    \caption{Qwen3}
    \label{fig:derp_qwen}
  \end{subfigure}%
  % --- prawy rysunek ---
  \begin{subfigure}[t]{0.52\textwidth}
    \centering
\includegraphics[width=\linewidth]{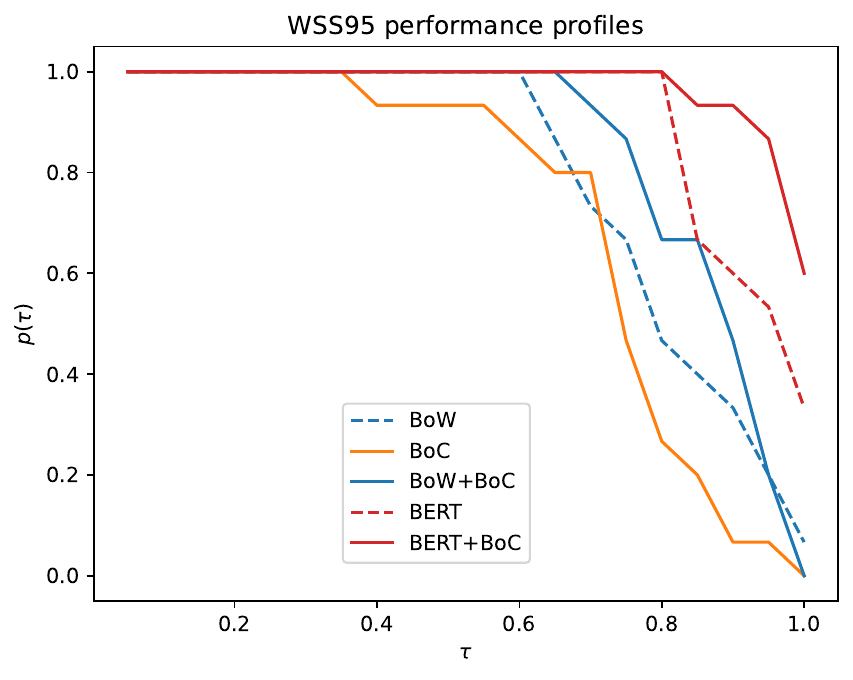}
    % \caption{NoW, BoC, BoW+BoC, BERT, BERT+BoC).}
    \caption{PubMedBERT}
    \label{fig:derp_bert}
  \end{subfigure}
  % \caption{Performance profiles on DERP based on SNOMED--CT + MeSH ontologies.}
  \caption{DERP performance profiles for neural representations with and without BoC augmentation. BoC uses MeSH--SNOMED CT vocabularies.}
  \label{fig:derp_qwen_bert}
\end{figure}

The performed experiments led us to the following observations:
\begin{description}
\item[1.] Based on the performance profiles for 15 medical case studies it can be concluded that the Qwen3+BoC representation outperformed other representations; it also achieved the best average performance among the evaluated representations, as measured by $WSS@95$---see Table \ref{tab:derp_representations} and Figure \ref{fig:derp_all}.
\item[2.] The performance profiles for the 15 DERP case studies indicate an overall improvement from concatenating each base representation (BoW, PubMedBERT, Jina, Gemma and Qwen3) with BoC; mean $WSS@95$ also increased for each base representation. Improvements occurred in 10 of 15 cases for BoW, PubMedBERT and Qwen3, and in 11 of 15 cases for Jina and Gemma---see Table \ref{tab:derp_representations} and Figures~\ref{fig:derp_all},~\ref{fig:derp_jina_gemma} and~\ref{fig:derp_qwen_bert}.
\item[3.] For the 15 DERP case studies the relative performance of the tested ontology combinations varied between cases---see Figure \ref{fig:ont1} and Table \ref{tab:derp_ontologies}.
\item[4.] The MedCAT annotator with the large dictionary based on the UMLS Metathesaurus and the tested NOBLE configurations showed case-dependent differences in screening performance (Figure \ref{fig:ont2} and Table \ref{T_wss_rf_cu_medcat_n}).
\end{description}

In light of observations 3) and 4), the majority of simulation experiments were conducted using BoC representations based on MeSH + SNOMED CT ontologies (which are most frequently used in the context of BoC representations). Observation 4) shows that MedCAT provides an alternative for generating BoC representations of medical texts. In practice, the choice of annotator depends on annotation performance in the target domain and the effort required to integrate it into the system supporting SLR processes.

%%%%---

\begin{tiny}
\begin{table*}
%\begin{table}[p]
\centering
\caption{Active learning $WSS@95$ metric values on DERP for the BoW and BoW+C text representations, for MeSH--SNOMED CT and NCI Thesaurus--ChEBI ontologies respectively ($\%$ denotes the percentage of screened articles at which the $WSS@95$ metric value was achieved).}
\label{tab:derp_ontologies}
\begin{tabular}{lllllllll}
		\hline
		Case name & BoW  & BoW  & BoW+C & BoW+C & BoW+C & BoW+C  \\ 
		& (\%) &  (WSS95) & (\%) & (WSS95) & (\%) & (WSS95) \\ \hline 
ACEInhibitors & 67 & 0.232 &  69 & 0.265  & 60 & {\bf 0.277} \\
ADHD & 29 & 0.659 &  27 & {\bf 0.683}  & 28 & 0.663 \\
Antihistamines & 84 & 0.111 & 78 & {\bf 0.153}  & 82 & 0.127 \\ 
AtypicalAntipsychotics & 82 & 0.108 & 81 & 0.130 & 54 & {\bf 0.140} \\ 
 BetaBlockers & 55 & 0.193 &  75 & 0.203  & 62 & {\bf 0.223} \\
CalciumChannelBlockers & 65 & 0.233 & 67& {\bf 0.245} & 63 & 0.212 \\ 
 Estrogens & 62 & 0.293 & 65 & {\bf 0.304} & 53 & 0.290 \\ 
NSAIDS & 49 & {\bf 0.382} & 56 & 0.351 & 60 & 0.342 \\ 
Opioids & 63 & 0.322 &  53 & {\bf 0.404} & 35 & 0.360 \\ 
OralHypoglycemics & 81 & 0.122 &  82 & {\bf 0.127} & 83 & 0.119 \\  
ProtonPumpInhibitors & 76 & 0.176 & 72 & 0.231 & 71 & {\bf 0.232} \\ 
SkeletalMuscleRelaxants & 54 & {\bf 0.320} & 51 & 0.314 & 56 & 0.293 \\ 
Statins & 68 & 0.258 & 72 & 0.233 & 67 & {\bf 0.261} \\ 
Triptans & 46 & {\bf 0.346} & 50 & 0.343 & 43 & 0.342 \\ 
UrinaryIncontinence & 63 & {\bf 0.260} & 60 & 0.242  & 60 & 0.218        
\end{tabular}
\end{table*}
\end{tiny}
%\end{table}
%%%%---
\begin{tiny}
\begin{table*}
%\begin{table}[p]
\centering
\caption{Active learning $WSS@95$ metric values on DERP for the BoW, BoC and BoW+C text representations, for the UMLS MedCAT annotator ($\%$ denotes the percentage of screened articles at which the $WSS@95$ metric value was achieved).}
\label{T_wss_rf_cu_medcat_n}
\begin{tabular}{lllllllll}
		\hline
		Case name & BoW  & BoW  & BoC & BoC & BoW+C & BoW+C  \\ 
		& (\%) &  (WSS95) & (\%) & (WSS95) & (\%) & (WSS95) \\ \hline 
ACEInhibitors & 67 & 0.232 & 51 & 0.236 & 66 & {\bf 0.292} \\ 
ADHD & 29 & 0.659 & 28 & 0.673 & 27 & {\bf 0.674} \\ 
Antihistamines & 84 & 0.111 & 77 & {\bf 0.182} & 80 & 0.146 \\ 
AtypicalAntipsychotics & 82 & 0.108 & 82 & 0.116 & 76 & {\bf 0.130} \\ 
BetaBlockers & 55 & 0.193 & 70 & 0.151 & 71 & {\bf 0.229} \\ 
CalciumChannelBlockers & 65 & {\bf 0.233} & 63 & 0.203 & 72 & 0.214 \\ 
Estrogens & 62 & 0.293 & 63 & {\bf 0.319} & 63 & 0.300 \\ 
NSAIDS & 49 & {\bf 0.382} & 57 & 0.349 & 54 & 0.377 \\ 
Opioids & 63 & 0.322 & 59 & 0.346 & 58 & {\bf 0.375} \\ 
OralHypoglycemics & 81 & {\bf 0.122} & 27 & 0.106 & 85 & 0.100 \\ 
ProtonPumpInhibitors & 76 & 0.176 & 75 & 0.190 & 73 & {\bf 0.222} \\ 
SkeletalMuscleRelaxants & 54 & 0.320 & 61 & 0.334 & 48 & {\bf 0.438} \\ 
Statins & 68 & 0.258 & 70 & 0.197 & 67 & {\bf 0.269} \\ 
Triptans & 46 & 0.346 & 49 & 0.345 & 45 & {\bf 0.352} \\ 
UrinaryIncontinence & 63 & {\bf 0.260} & 48 & 0.206 & 58 & 0.192       
\end{tabular}
\end{table*}
\end{tiny}
%\end{table}
%%%%---
%%%%%%%%%%%%%%%%%%%%%%%%%%%%%%%%%%%%%%%%%%%%%%%%%%%
%\begin{tiny}
%\begin{table*}
\begin{sidewaystable} 
\begin{tiny}
%\begin{table}[p]
\centering
\caption{$WSS@95$ (denoted as $WSS$) metric values on DERP for the BoW, BoW+C, BERT, BERT+BoC, Jina, Jina+BoC, Gemma, Gemma+BoC, Qwen3 and Qwen3+BoC  text representations, active learning with cluster initialization and uncertainty sampling, the RF algorithm and MeSH--SNOMED CT ontologies ($\%$ denotes the percentage of screened articles at which the $WSS@95$ metric value was achieved).}
\label{tab:derp_representations}
\begin{tabular}{lllllllllllllllllllllll}
		\hline
		Case name & BoW  & BoW  & BoC & BoC & BoW & BoW & BERT & BERT & BERT &  BERT & Jina & Jina & Jina & Jina & Gemma & Gemma & Gemma & Gemma & Qwen3 & Qwen3 & Qwen3 & Qwen3 \\ 
        & & & & &+C & + C & &  & +BoC & +BoC & & & +BoC & +BoC & & & +BoC & +BoC & & & +BoC & +BoC  \\  
		& (\%) &  (WSS) & (\%) & (WSS) & (\%) & (WSS) & (\%) &  (WSS) & (\%) & (WSS) & (\%) & (WSS) & (\%) &  (WSS) & (\%) &  (WSS) & (\%) & (WSS) & (\%) & (WSS) & (\%) &  (WSS)\\ \hline 
ACEInhibitors  & 67 & 0.232 & 71 & 0.213 & 69 & 0.265 & 65 & 0.303 & 66 & 0.294  & 64 & 0.313 & 59 & 0.308  & 61 & 0.311 & 64 & 0.299 & 56 & {\bf 0.384} & 60 & 0.353                \\ 
ADHD & 29 & 0.659 & 27 & 0.676 & 27 & 0.683 & 24 & 0.694 & 27 & 0.678   & 23 & 0.715 & 24 & 0.713    & 25 & 0.693 & 24 & 0.707    & 21 & {\bf 0.737} & 22 & 0.729    \\ 
Antihistamines & 84 & 0.111 & 77 & 0.143 & 78 & 0.153 & 77 & 0.142 & 73 & 0.164  & 71 & 0.234 & 71 & 0.238     & 71 & {\bf 0.238} & 71 & 0.221  & 71 & 0.235 & 70 & 0.226      \\ 
AtypicalAntipsychotics & 82 & 0.108 & 79 & 0.139 & 81 & 0.130 & 69 & 0.179 & 69 & 0.184        & 75 & 0.157 & 60 & 0.162   & 77 & 0.154 & 75 & 0.174  & 70 & 0.179 & 74 & {\bf 0.190} \\ 
BetaBlockers & 55 & 0.193 & 78 & 0.168 & 75 & 0.203 & 71 & 0.240 & 64 & 0.261       & 69 & 0.250 & 64 & 0.294  & 70 & 0.255 & 68 & 0.264  & 61 & 0.267 & 64 & {\bf 0.304} \\ 
CalciumChannelBlockers & 65 & 0.233 & 66 & 0.211 & 67 & 0.245 & 71 & 0.207 & 69 & {\bf 0.246}       & 74 & 0.201 & 56 & 0.212  & 72 & 0.186 & 70 & 0.232  & 73 & 0.207 & 68 & 0.240\\ 
Estrogens& 62 & 0.293 & 67 & 0.281 & 65 & 0.304 & 54 & 0.396 & 50 & 0.390  & 61 & 0.341 & 59 & 0.363 & 56 & 0.387 & 60 & 0.349 & 54 & 0.400 & 54 & {\bf 0.404} \\ 
NSAIDS & 49 & 0.382 & 39 & 0.287 & 55 & 0.351 & 60 & 0.324 & 59 & 0.346 & 56 & 0.383 & 53 & 0.395 & 70 & 0.250 & 61 & 0.332 & 55 & 0.358 & 53 & {\bf 0.396}\\ 
Opioids & 63 & 0.322 & 72 & 0.175 & 53 & 0.404 & 47 & 0.453 & 50 & 0.380 & 52 & 0.425 & 43 & 0.435 & 39 & 0.420 & 52 & 0.435  & 38 & 0.438 & 25 & {\bf 0.458} \\ 
OralHypoglycemics & 81 & 0.122  & 84 & 0.101 & 82 & 0.127 & 83 & 0.116 & 81 & 0.138 & 82 & 0.125 & 80 & {\bf 0.146} & 83 & 0.116 & 64 & 0.120 & 73 & 0.122 & 77 & 0.135\\ 
ProtonPumpInhibitors & 76 & 0.176  & 77 & 0.179 & 72 & 0.231 & 75 & 0.202 & 71 & {\bf 0.244}     & 68 & 0.235 & 68 & 0.229  & 73 & 0.204 & 71 & 0.213 & 69 & 0.242 & 67 & 0.238  \\ 
SkeletalMuscleRelaxants & 54 & 0.320  & 61 & 0.247 & 51 & 0.314 & 60 & 0.354 & 51 & {\bf 0.421}  & 45 & 0.385 & 63 & 0.294 & 53 & 0.417 & 62 & 0.300   & 60 & 0.327 & 36 & 0.399 \\ 
Statins & 68 & 0.258 & 64 & 0.267 & 72 & 0.233 & 62 & 0.329 & 60 & 0.346  & 72 & 0.227 & 57 & 0.345 & 64 & 0.229 & 58 & 0.356 & 70 & 0.250 & 52 & {\bf 0.409}\\ 
Triptans & 46 & 0.346 & 61 & 0.278 & 50 & 0.343 & 46 & 0.343 & 51 & 0.336 & 55 & 0.298 & 51 & 0.341 & 67 & 0.252 & 56 & 0.310 & 60 & 0.301 & 49 & {\bf 0.350}   \\ 
UrinaryIncontinence & 63 & 0.260 & 67 & 0.191 & 60 & 0.242 & 60 & 0.257 & 69 & 0.260  & 64 & 0.263 & 65 & 0.293 & 69 & 0.264 & 62 & 0.296 & 64 & {\bf 0.301} & 56 & 0.294\\ 

\end{tabular}
%\end{table*}
\end{tiny}
\end{sidewaystable} 
%\end{tiny}

\FloatBarrier

\section{Conclusions}
%%%%---
This article presents a systematic experimental analysis of the use of BoC representations in systems supporting SLR processes based on machine learning. Building on the framework studied in \cite{Pytlak_2025}, we evaluate the augmentation of neural text representations with BoC features, compare selected ontology and annotator configurations.

The conducted experiments not only enhance the state of knowledge related to the creation of BoC representations but also lead to conclusions that can be used in the selection of embeddings applied for document classification. In particular, the results obtained with different annotators whose dictionaries were built based on different ontologies indicate that combining text embeddings and BoC features may lead to better results in the active learning process supporting SLRs than either representation used separately. On DERP, mean workload savings increase for each tested base representation, with Qwen3+BoC achieving the best average performance. 

More work is needed to verify the utility of the BoC representation for SLRs in other domains as well as to compare the performance of conventional machine learning algorithms with such semantically augmented text representations to that achieved by large language models used directly as classifiers. Further work could examine how concept information, feature dimensionality and query selection contribute to the observed gains. It may be particularly interesting to investigate possible hybrid methods combining the strengths of these two approaches and hopefully eliminating some of their limitations.

\bibliographystyle{unsrt}  

\bibliography{references}
%\bibliographystyle{splncs04}

%%\appendix
%%\section{Appendix}
%%You may include other additional sections here.

\end{document}